\documentclass[10pt,letterpaper]{article}
\usepackage[top=0.85in,left=2.75in,footskip=0.75in]{geometry}

\usepackage{amsmath,amssymb}
\usepackage{url}
\usepackage{changepage}

\usepackage{textcomp,marvosym}

\usepackage{cite}
\usepackage{tabularx}

\usepackage[nopatch=eqnum]{microtype}
\DisableLigatures[f]{encoding = *, family = * }

\usepackage[table]{xcolor}

\usepackage{array}
\usepackage{caption}
\usepackage{subcaption}

\newcolumntype{+}{!{\vrule width 2pt}}

\newlength\savedwidth

\raggedright
\usepackage[aboveskip=1pt,labelfont=bf,labelsep=period,justification=raggedright,singlelinecheck=off]{caption}

\makeatletter
\renewcommand{\@biblabel}[1]{\quad#1.}
\makeatother

\usepackage{lastpage,fancyhdr,graphicx}
\usepackage{epstopdf}
\fancyheadoffset[L]{2.25in}
\fancyfootoffset[L]{2.25in}
\begin{document}
\vspace*{0.2in}

\begin{flushleft}
{\Large
\textbf\newline{Tracking behavioural differences across chronotypes: A case study in Finland using Oura rings} 
}
\newline
\\
Chandreyee Roy *,
Kunal Bhattacharya, and
Kimmo Kaski

\bigskip
Department of Computer Science, Aalto University School of Science, Espoo, Finland

\bigskip

%
%





* chandreyee.roy@aalto.fi

\end{flushleft}
\section*{Abstract}

Non-invasive mobile wearables like fitness trackers, smartwatches and rings allow for an easier and relatively less expensive approach to study everyday human behaviour when compared to traditional longitudinal methods. Here we have utilised smart rings manufactured by Oura to obtain granular data from nineteen healthy participants over the time span of one year (October 2023 - September 2024) along with monthly surveys for nine months to track their subjective stress during the study. We have investigated longitudinal sleep and activity patterns of three chronotype groups of participating individuals: morning type (MT), neither type (NT) and evening type (ET). We find that while ET individuals do not seem to lead as healthy life as the MT or NT individuals in terms of overall sleep and activity, they seem to have significantly improved their habits during the duration of the study. The activity in all chronotype groups varies across the year with ET showing an increasing trend. Furthermore, we also show that the Daylight Saving Time changes affect the MT and ET chronotypes, oppositely. Finally, using a mixed-effects regression model, we show that an individual's perceived stress is significantly associated with their time spent in bed during the night time sleep, monthly survey response time, and chronotype, while accounting for individual variability. 
\section*{Author summary}
Consumer-grade wearables like rings and smartwatches are a non-invasive way of monitoring daily health and activity remotely and they provide researchers with real-time granular data in naturalistic settings. To understand the long-term behaviour of the three chronotype groups: morning, evening, and neither type, we used data collected from Oura rings for one year. 
We present seasonal changes in the sleep and activity patterns of all participants, along with the effects of time changes during autumn and spring daylight saving time on the sleep quality of these groups. Furthermore, we used mixed effects models to predict the perceived stress of the individuals which was tracked monthly. We demonstrated that evening types are able to improve their sleep and activity over time and while the reported stress is non-linear in time, the neither types responded to having significantly lesser stress than the other groups. 



\section*{Introduction}
In the past few years, there has been a rapid growth of wearable mobile devices such as fitness trackers and smart watches that are being used by millions of people around the world \cite{huhn2022impact, kazanskiy2024review}. This phenomenon 
started with just simple trackers that were able to track the steps and basic activity of the user, but by now they have been developed to measure and collect data of the user's 
sleep, heart rate, heart rate variability, calories burned, daily activity, menstrual cycles, among others \cite{lee2023accuracy}. The added advantage of this personal data is that it can be synchronized to their respective applications on users' smartphones to provide them with in-depth analysis of their own health and progress reports. 
There are several motivations for the widespread adoption of these devices \cite{perez2019large}. One of them is that 
users are driven by the need to improve their physical fitness and these devices can serve as aid by providing instant feedback along with features such as reminders to move, to achieve daily step goals and details of the workout \cite{sitra}. Additionally, they have already proven quite useful in tracking the early onset of infectious diseases \cite{mishra2020pre, grzesiak2021assessment,li2017digital,mayer2022consumer }. 

Several users are also motivated by the need to improve their sleep due to the ability of these devices to gauge an individual's sleep quality throughout the night \cite{kinnunen2020feasible, alzueta2022tracking}. Sleep has often been considered one of the pillars of health, along with nutrition and exercise, due to its profound impact on all aspects of our daily functioning \cite{walker2017we}. Several studies have shown that neglecting sleep can lead to a cascading number of negative health outcomes \cite{durmer2005neurocognitive,knutson2007metabolic}, and therefore prioritizing it is emphasized to improve one’s quality of life. Since sleep plays such an important role in human physiological processes, cognition, and overall wellbeing \cite{grandner2022sleep}, its monitoring and assessment have taken precedence among those who want to lead a healthy life.   

Traditionally, an individual's sleep and health have been assessed using polysomnography (PSG) \cite{jafari2010polysomnography} and actigraphy \cite{fekedulegn2020actigraphy}, but 
they have mainly been used for research purposes only. The advent of advanced fitness trackers from all price brackets have allowed humans to conveniently assess their sleep and daily activities on a regular basis. These devices encompass a wide range of gadgets, such as watches, rings, and activity bands, among others that are worn close to or on the body \cite{binkley2003predicting,ko2022quantification}. These devices typically have sensors that can collect their users' physiological data, which are then processed and interpreted using algorithms and machine learning methods by health applications that can nowadays be downloaded easily on smartphones. This method of monitoring one's health is now very popular among the general population \cite{sitra}.

Conventional methods for detecting sleep and activity generally involve sleep laboratories \cite{portier2000evaluation,bruyneel2011sleep} and
the use of a myriad of wires, electrodes, and overnight stays by subjects, which does not simulate a natural environment for studying sleep for extended periods of time. These wearables have provided a way in which we can now capture sleep patterns in the real world without being too intrusive \cite{leeder2012sleep, ko2015consumer}. Today, the most popular method is to use wristwatches that detect movement using accelerometers and rings that can measure heart rate variability and body temperatures \cite{shelgikar2016sleep}. The wearables offer an unparalleled insight into one’s sleep cycles, sleep quality, and even circadian rhythm \cite{koskimaki2018we}.  

In the current study, we have used Oura rings \cite{kinnunen20180061} to investigate behavioural health patterns among volunteers for a period of one year. Some of the sleep and activity metrics of the Oura rings have been previously validated against gold standard methods and therefore considered as a suitable consumer grade device for the present study \cite{kristiansson2023validation, svensson2024validity}. Our main objective is to study the differences in circadian patterns of a group of volunteers from a long-term perspective. Here, we focus on the circadian rhythms of volunteers, as it plays a pivotal role in their sleep-wake cycles, eating patterns, hormone cycles, and other physiological processes \cite{foster2020sleep}. Disruptions in individuals' circadian rhythms due to lifestyle, travel, and work shifts can have a major impact on their health and well-being \cite{merikanto2013evening}.
We have identified each individual's chronotype \cite{roenneberg2015having} his or her sleeping or waking time preferences using 
the Horne and Ostberg morningness eveningness questionnaire \cite{horne1976self}.  
Consequently, if a person is unable to get up in the morning but gets most of the daily tasks done during the latter part of the day, then his or her chronotype would resemble that of an “owl” and is more commonly known as the evening type (ET) \cite{roenneberg2007epidemiology}.  
On the other hand, if one prefers getting up early in the morning and starts to feel groggy in the evening, then the chronotype would resemble that of a “lark” (also known as the morning type (MT)). Most people in a large population fall within the spectrum of these two categories of larks and owls. Those individuals in the middle of the spectrum constitute the majority of the people and are known as "third birds" of neither types (NT) \cite{roy2021morningness}. 

While the proverb “early to bed, early to rise makes a man healthy, wealthy, and wise” suggests that larks lead better lives than owls, there are some studies showing that owls can also lead healthy lives if they follow their own sleeping patterns \cite{salfi2022fall}. This is because alignment of one's body clock with their inherent chronotypes, results in most productivity in their daily lives. Unfortunately, misalignment of one's chronotype is quite common \cite{chellappa2020circadian} even though it is considered an issue that leads to irregular sleep patterns, unhealthy eating habits, and general poor health. Most of the time people are simply unaware of this misalignment or are unable to follow their inherent circadian rhythms due to non-flexible work or school schedules \cite{van2021acute}. This is predominately seen to affect ET individuals\cite{merikanto2022evening, makarem2020evening} due to the global bias towards morning friendly schedules, and therefore it is the most vulnerable group among the chronotypes. In the present study, our aim is 
to use behavioural data of individuals related to their sleep and activity patterns 
collected from Oura rings to identify the long-term differences between chronotypes. 

The present study was conducted among participants living in the Helsinki capital region, Finland, which is situated at northern lattitude (60.2° N). 
The contrast in sunlight hours at this latitude, along with the weather conditions, provides an interesting backdrop for studying the behaviour of chronotypes throughout the year. In the present study we find that while ET individuals have poorer sleep quality in general, 
they can potentially improve with time.
We also find that the Daylight saving times (DST) in autumn and spring affect the chronotypes differently. 
In addition, the time spent in bed for night sleep was significantly associated with less reported stress. Compared to MTs, the NT individuals reported significantly lower levels of stress, while no significant differences were found between MTs and ETs.


\begin{figure}[h]
    \centering
    \includegraphics[width=\linewidth]{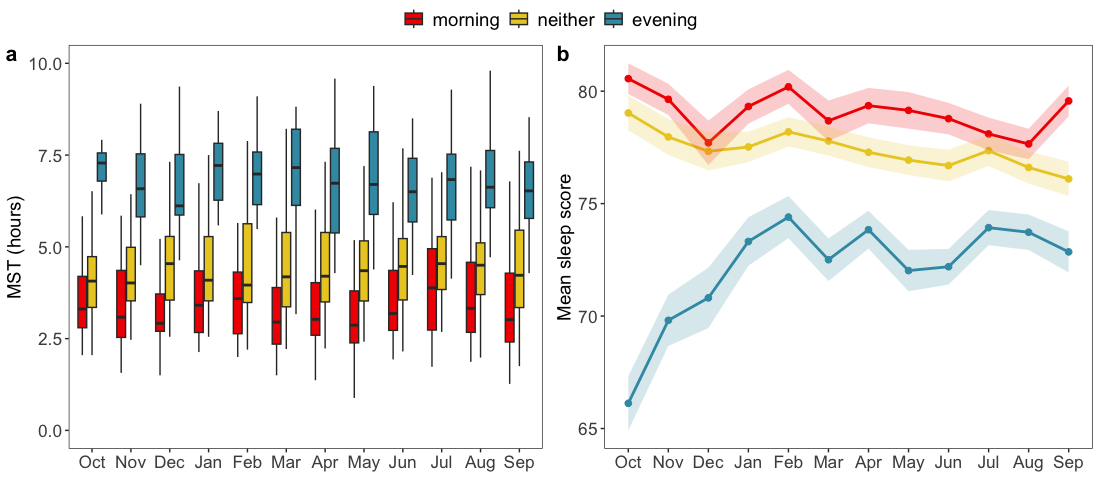 }
    \caption{\textbf{Mid sleep time (MST) and sleep scores of the chronotypes.} (a) Evening-types (ET) in blue have later MSTs while morning-types (MT) in red have earlier MSTs when compared with neither-types (NT) in yellow. 
    We observe seasonal changes in all chronotypes for winter and summer holidays, showing later MSTs as can be seen from Fig. \ref{fig:sesonal_var} in the SI. (b) The sleep scores of the Oura ring takes into account several different sleep metrics for scoring as displayed here. ETs are observed to have poorer sleep scores when compared to other chronotypes, but their scores improve considerably with time. The shaded region around the line plots show the 95$\%$ confidence interval of the data. }  
    \label{fig:mid_sleep_scores}
\end{figure}

\begin{figure}[h]
    \centering
    \includegraphics[width=\linewidth]{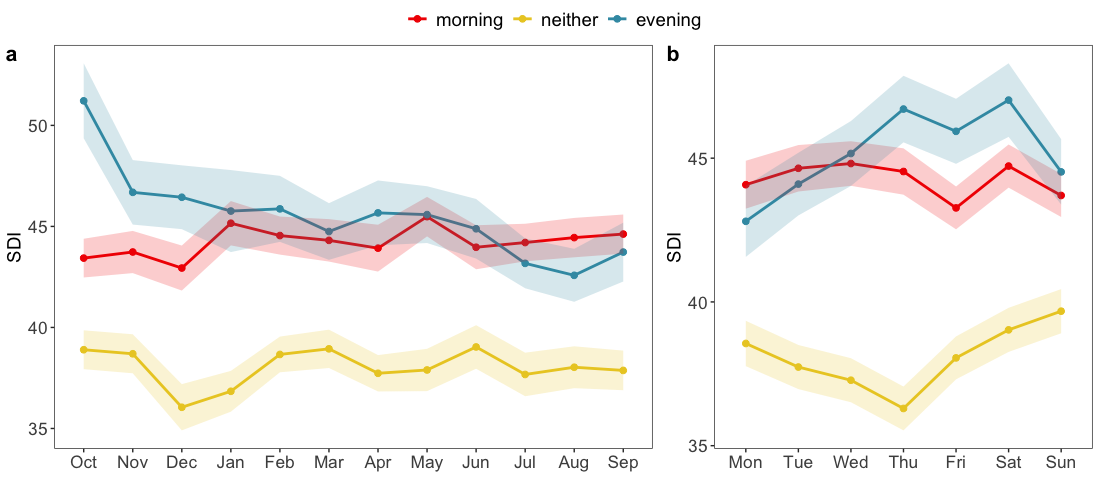}
    \caption{\textbf{Sleep diversity index (SDI).} The SDI is calculated using Shannon entropy from Eq.\ref{eq:shannon_entropy} and \ref{eq:sdi} utilising sleep hypnograms (see Fig. \ref{fig:sleep_hypnogram}) such that the number ranges from 1 to 100. Higher values indicate a more fragmented sleep during the night time. We find that NTs display lowermost values throughout the (a) year, as well as (b) on a weekly basis indicating better sleep quality than both MTs and ETs. The shaded regions represent $95\%$ confidence interval. The ETs had poorer values of SDI at the start of the study, but improved with time. We also observe that the ETs regularly have poorer quality sleep towards the later part of the week. } 
    \label{fig:sdi}
\end{figure}
\section*{Results}
\subsection*{Differences in sleeping patterns among chronotypes}
\begin{figure}[h]
    \centering
\includegraphics[width=0.7\linewidth]{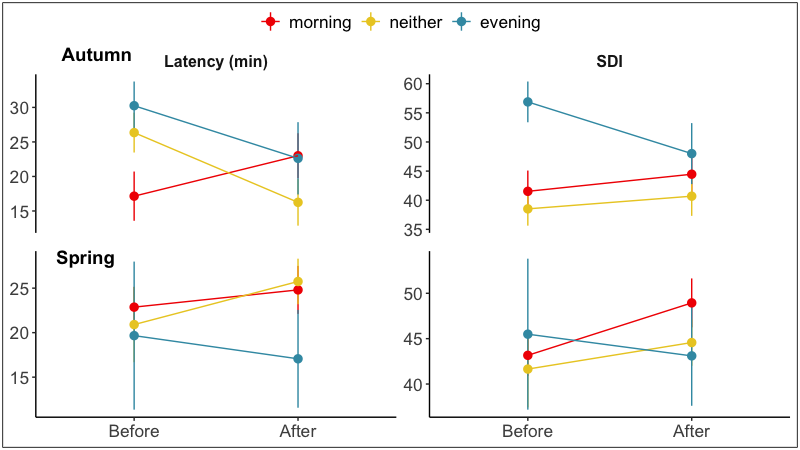}
    \caption{\textbf{Daylight saving time. }Differences in SDI and sleep latency among the chronotypes before and after the time is changed in autumn and in spring. In autumn, the latency of the ETs decreases and that of the MTs, increases. This is also observed for SDI values before and after the DST change in the autumn. 
    We do not find much differences in the latency during spring DST change, but, the SDI of MTs increases again during this time. The most significant changes ($p<0.05$) occur in sleep latency for MTs and NTs in the autumn DST.  }
    \label{fig:dst}
\end{figure}

\begin{figure}[h]
    \centering
    \includegraphics[width=\linewidth]{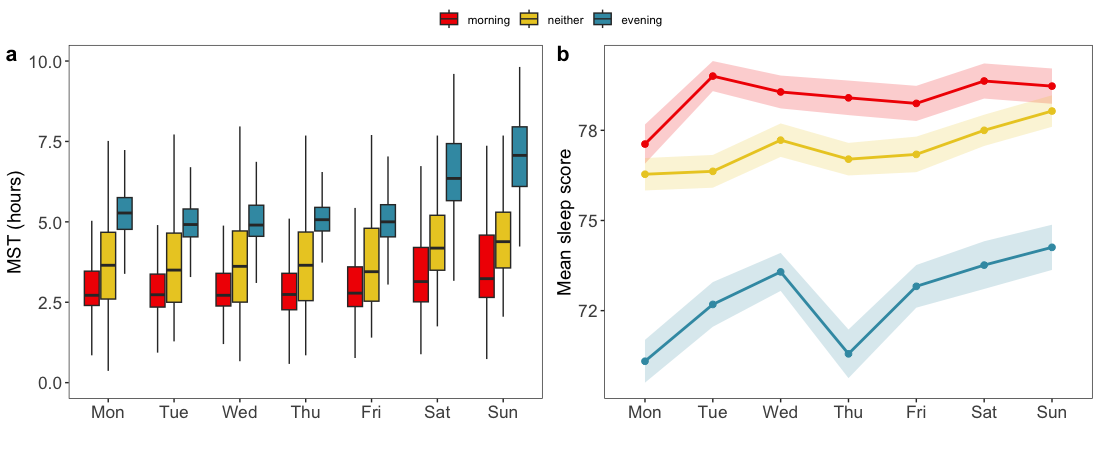}
    \caption{\textbf{Mid sleep times on weekly basis.} The MSTs of all chronotypes follow different patterns on weekdays and weekends. (a) On weekends the MST of all chronotypes are later than on weekdays. (b) The sleep scores also show and increasing trend towards the weekends.    }
    \label{fig:mid_point_weekly}
\end{figure}

\begin{figure}[h]
    \centering
    \includegraphics[width=0.6\linewidth]{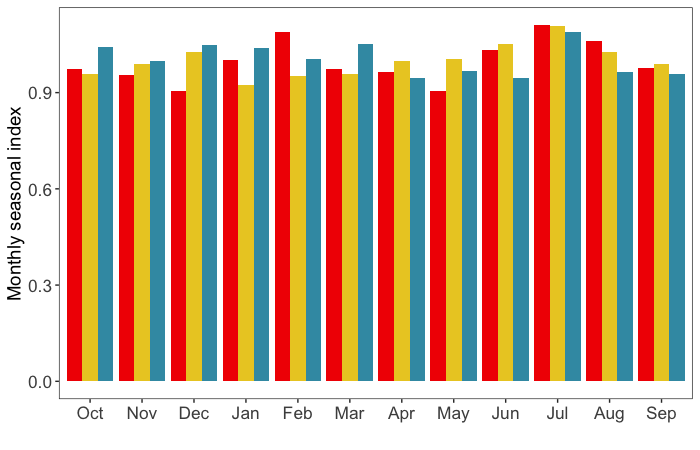}
    \caption{\textbf{Monthly seasonal indices of the MSTs across chronotypes.} A bar chart showing the seasonal indices for each month to show how much the values deviate from the overall average values in each group. The seasonal indices are calculated as monthly mean/overall group mean. A clear peak can be seen during summer time indicating that all the participants slept at later hours around that time. }
    \label{fig:sesonal_var}
\end{figure}

In the present study we first observe the known differences in the sleeping patterns of the participants' chronotypes. 
The mid-sleep time (MST) for all chronotypes on free days is depicted in Fig. \ref{fig:mid_sleep_scores}(a). It is well known that 
humans tend to follow their own circadian rhythms on free days 
due to experiencing less effect 
of work or study-related schedules \cite{roy2021morningness} 
(See also the weekly variation of the MSTs in Fig. \ref{fig:mid_point_weekly}(a) in the SI). A similar pattern is also observed for work days, but is not shown in the present manuscript. The MSTs for ET individuals are observed to be significantly later in the study period compared to MT or NT. A seasonal variation is also observed for all chronotypes with individuals having late MSTs during winter and summer holidays, as can be seen from the seasonal indices shown in Fig. \ref{fig:sesonal_var} in the SI. 
\begin{figure}[h]
    \centering
    \includegraphics[width=\linewidth,height=6cm]{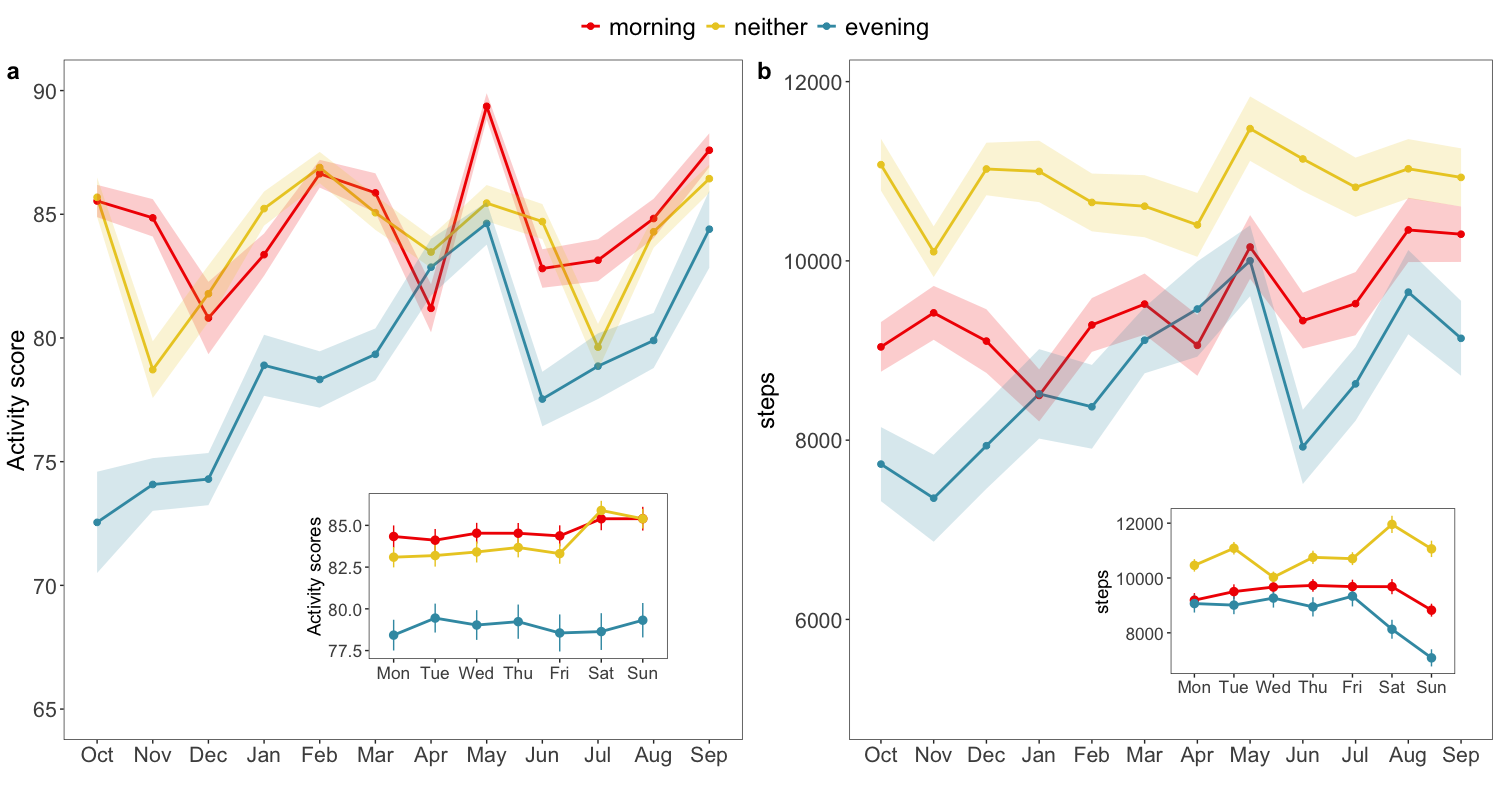}
    
    \caption{\textbf{Activity scores and steps.} The monthly variation of the activity scores throughout the study period is shown in (a) along with the weekly variation in the inset of the figure. The ETs have low scores in the beginning but their scores are observed to have increased with time. Peaks in activity scores of the MTs and ETs coincide with the annual summer and winter activity schedules of Finland. A similar trend is observed in the average monthly steps taken by the participants as shown in Fig. (b). NTs have have maintained higher number steps when compared with other chronotypes and MTs and ETs are observed to have improved their average of steps by the end of the study.   }
    \label{fig:activity}
\end{figure}

\begin{figure}[h]
    \centering
    \includegraphics[width=\linewidth]{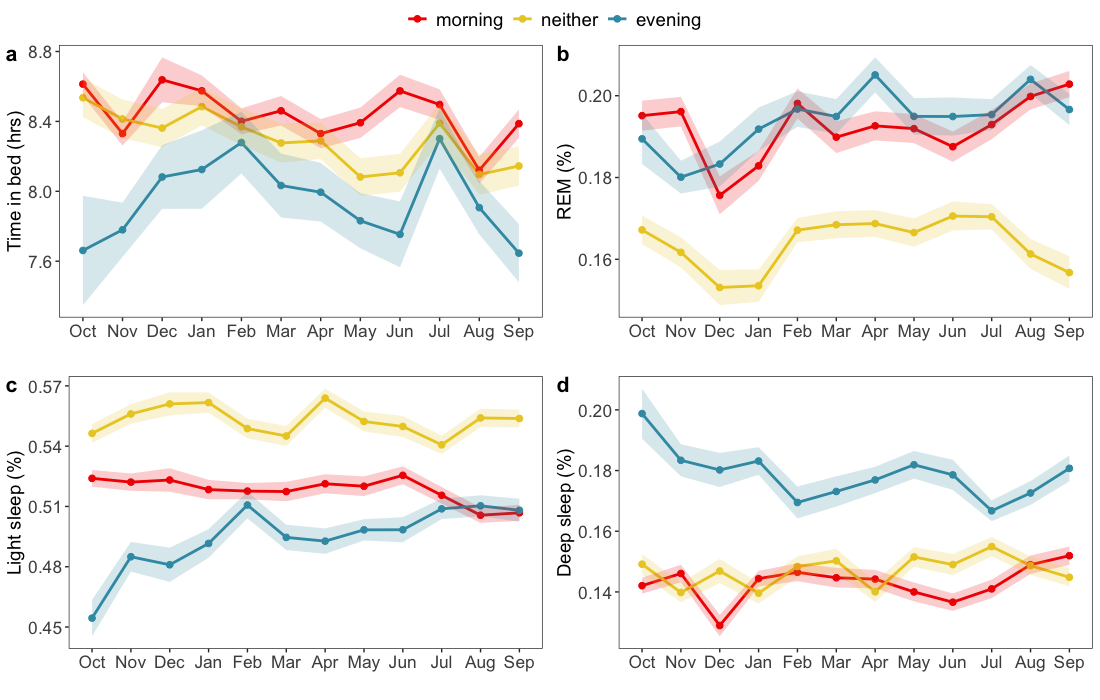}
    \caption{\textbf{Some of the sleep metrics that are detected by Oura rings and used to calculate the sleep scores.} In (a) we have shown the time spent in bed during night sleep events. We observe double peaks during winter and summer holiday times with a decresing trend during the spring time. In (b), (c) and (d) the percentage of night sleep time spent in REM, light and deep sleep stages out of the total sleep duration respectively are displayed.  }
    \label{fig:sleep_stage}
\end{figure}

The Oura rings assign a sleep score for each sleep event taking 
into account several sleep metrics detected by the rings (see Materials and Methods section). 
In Fig. \ref{fig:mid_sleep_scores}(b) we have shown
the average monthly sleep scores for the different chronotypes. Although we observed the worst sleep scores for ETs compared to other chronotypes, they were found to make significant progress in their sleep scores during the course of the study.
Some of the sleep metrics used to measure sleep scores are 
plotted in the SI in Fig. \ref{fig:sleep_stage}.

\begin{figure}[h]
    \centering
    \includegraphics[width=\linewidth]{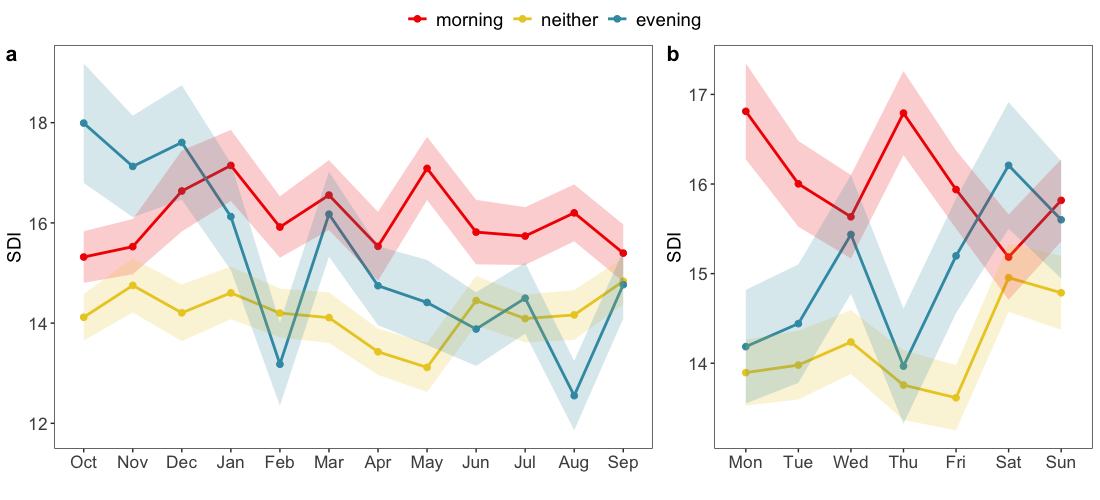}
    \caption{\textbf{SDI for 20 min sampling frequency.} The SDIs calculated from a shorter epoch has more variation but overall shows similar trends as the 40 minute epoch. (a) The ET had an improving sleep quality over the months and the MT and ET had stable values. (b) On a weekly basis, the ETs and the MTs observe higher SDI towards end of the week, while MTs remain stable overall.  }
    \label{fig:sdi_20}
\end{figure}

Next, we calculate the sleep diversity index (SDI) that uses Shannon entropy to capture the diversity of sleep stages, thus quantifying sleep disruptions. It is found to have a positive correlation with the standard measurements, such as the sleep fragmentation index as seen in Ref. \cite{suppini2009sleep}. In the present study, we have used the sleep hypnogram recorded by Oura to calculate the SDI for each of the sleep sessions. (See Materials and Methods for calculations). In Fig. \ref{fig:sdi}(a) we have plotted the monthly variation of the SDI. A higher value of SDI suggests more disruptive sleep and we observe that NTs have better quality sleep throughout the year compared to both ETs and MTs. Additionally, ETs have better SDI values towards the latter part of the study. In the weekly variation in Fig. \ref{fig:sdi}(b) we observe that sleep quality for ETs and NTs is more disruptive toward the end of the week, while MTs are able to maintain their sleep quality. Since each data point in the hypnogram is five minutes, we considered a sampling frequency of 40 minutes for the SDI calculation, which ensures a maximum of eight data points within each sampling. 
We also carried out the same analysis using a smaller sampling frequency of 20 minutes and is shown in Fig. \ref{fig:sdi_20} in the SI. In the latter case, we observe similar trends but higher fluctuations in the values.

We have also investigated 
the SDI and sleep latency during daylight saving time (DST) changes both in the autumn (29 October, 2023) and spring (31 March, 2024) times. Since the changes occurred on Sunday and to exclude the effects of the weekend on sleeping patterns, we consider the average of nights from Monday to Thursday before and after the change in DST as shown in Fig. \ref{fig:dst}. If any of the volunteers were found travelling outside of Europe around this time, then they were not included in the DST calculations. 
We observe that during autumn the latency for ET and NT decreases but for MT increases whereas not much difference can be found in spring. We also find that the SDI lowers for ET in autumn, but increases for MT in spring. The changes in sleep latency in autumn for NTs and MTs were found to be significant ($p<0.05$) from paired t-test calculations. 

\subsection*{Activity differences among the chronotypes}
The monthly variation in the activity scores recorded by the Oura ring for each day is plotted in Fig. \ref{fig:activity}(a). We find that while ETs scored low values at the beginning of the study, they increased with time while the other chronotypes maintained steady scores. We observe mainly two peaks for NTs and MTs during February-March and during the month of May. 
A significant jump in sleep scores was observed in January for ETs. The weekly variation is shown in the inset of the figure and we observe that ETs and MTs have better scores on weekends than on weekdays. The total number of steps taken by participants each day and its monthly variation is plotted in Fig. \ref{fig:activity}(b). The weekly variation of the steps are shown in the inset and all chronotypes are observed to have low steps on the weekends compared to the weekdays. NTs have maintained the highest step counts throughout the duration of the study and MTs and ETs have increased the average monthly step counts over time. Here we also observe peaks in the average steps 
for NTs and MTs and an overall increasing trend for ETs till May. The summer months of June and July are observed to be a period of low activity for the volunteers in the study.

\subsection*{Subjective stress from questionnaires}
We conducted a total of nine monthly questionnaires from January, 2024 till September, 2024 
and tracked the subjective perception of stress. 
The coefficients of the models discussed in the Methods section with comparison statistics are listed in Table \ref{tab:model}. Based on likelihood ratio tests, the final model has the best fit and also the lowest AIC. Adding the age, gender, BMI or step counts did not improve the model and therefore has not been included in the present paper. We find that the survey response time, which was broadly speaking answered on a monthly basis, has a significant non-linear effect on stress. 
We also find that the NTs have significantly lower stress than the reference group (MT), but the ETs do not differ significantly from them. Spending more time in bed in the weekdays is associated with lower stress level. The random intercepts account for the individual variability in the stress.   

\section*{Discussion}
In this study, we have used Oura rings to track nineteen individuals for one year and observed monthly and weekly variations in their sleep and activity pattern. Through our preliminary results of sleep and activity data, we have found some distinct seasonal patterns across all chronotypes. Although the study contained healthy participants with no known disease diagnosed, ETs were found to have sleep and activity metrics that are not as good as the other groups (MT and NT). However, during the duration of the study, the ETs have made significant improvements in their sleeping and activity habits. While they continued sleeping at later hours, their sleep scores increased considerably in the middle of the study along with the total duration of sleep. We conducted a final questionnaire at the end of the study period regarding the individual experiences, answered by fourteen out of the nineteen participants. When asked whether they were able to ``follow their natural circadian rhythms"  by adjusting their schedules and activities to match their inherent rhythms, all of them answered either ``sometimes" or ``often". This sentiment is also observed from the data. We observe that their sleep quality has also improved with time. 

We find similar improvements in the activities of ETs and a noticeable jump in January may indicate the positive impacts of New Year resolutions for ETs. Interestingly, the double peaks in the activity scores are indicative of various sports activities that are played around this time in Finland. Finland has a short skiing holiday in February or March and people also participate in several outdoor activities during the end of spring. Similar trends are also observed in the average monthly steps taken by the participants. A noticeable fall of activity scores and steps during the summer holiday months (June and July) for all chronotype groups could be an indication of a popular aspect of Finnish culture in which individuals visit their summer cottages with their families. 
In addition, we find that a longer time spent in bed for night sleep cause less stress among participants and that 
NTs report lesser stress than the other groups. Although our model 
shows a non-linear relationship of stress with time, it should be noted that there may be several factors that directly or indirectly affect the feeling of stress. For this reason, it may be difficult to capture the confounding factors causing stress with just one question. In the future, we plan to use standard stress related questionnaires along with objective measurements.

We have also studied how the chronotypes are affected by the DST changes, especially in the context of Finland, since the sunlight hours change drastically around this time of the year. In the spring the clock is turned forward by one hour and in autumn the clock is turned back by one hour. This change is known to cause disruption to individual's sleep habits \cite{medina2015adverse,lahti2008transitions, harrison2013impact} which may eventually cause some health issues. In our study, we find that while MTs and ETs experience opposite effects of the change, the most significant impact occurs in autumn for sleep latency. During this change, NTs show significantly lower sleep latency, whereas MTs experience an increase just after the change. The ETs may benefit from the extra hour of sleep as seen by the lowering of latency and SDI although these changes are not statistically significant.
In the spring-time DST change we do not observe any major changes in the latency but the SDI increase in MTs is an indication of a more disruptive sleep following the change. In future studies, we plan to use more health metrics collected by the rings and other devices to provide a more comprehensive picture of the vulnerabilities faced by the chronotypes around DST. 

\begin{table}[h]
\begin{tabular}{l|l|l|l}
\hline
\hline
                      & \textbf{Null Model}  & \textbf{Intermediate model} & \textbf{Final Model }    \\
                      \hline
Random effects        & \multicolumn{3}{c}{Variance (SD)}                  \\
\hline
Participant intercept & 0.38 (0.61) & 0.39 (0.63)        & 0.37 (0.61)     \\
Residual              & 0.64 (0.8)  & 0.56 (0.75)        & 0.54 (0.74)     \\
\hline
Fixed effects         & \multicolumn{3}{c}{Coefficients (SE)}              \\
\hline
$Z_1(Time)$              &             & -2.40 (0.76) **    & -2.24 (0.75) ** \\
$Z_2(Time)$            &             & 2.33 (0.76) **     & 2.25 (0.75) **  \\
Neither type$^a$         &             &                    & -0.80 (0.35) *  \\
Evening type$^a$         &             &                    & -0.23 (0.43)    \\
Time in bed (Weekday) &             &                    & -0.21 (0.10) *  \\
Time in bed (Weekend) &             &                    & 0.10 (0.09)     \\
\hline
\hline
\multicolumn{4}{l}{Model fit statistics}                                   \\
\hline
AIC                   & 410.92      & 396.36             & 394.33          \\
BIC                   & 420.07      & 411.61             & 421.77          \\
Log Likelihood        & -202.46     & -193.18            & -188.16     \\
ICC (adjusted)       & 0.37     & 0.41            & 0.41     \\
ICC (unadjusted)        & 0.37     & 0.38          & 0.33     \\
\hline
\end{tabular}
$^a$ Reference level: Morning type \\
SD: Standard deviation \\
SE: Standard error \\
ICC: Interclass correlation coefficient \\
Significance levels: $p<0$ (***), $p<0.001$ (**), $p<0.01$ (*)
\caption{\textbf{Summary of the coefficients of the Linear Mixed Effects models for predicting stress.} The Time variable here refers to the nine time points when the questionnaires were answered by the participants. $Z_1(Time)$ and $Z_2(Time)$ capture the non-linear trends of stress with time and has a significant effect. MTs have higher stress when compared to NTs but did not differ significantly from the ETs and more time spent in bed on the weekdays cause less stress among the individuals. The models utilise individual variation in form of random intercepts of stress. The correlations between the fixed effects is displayed in Table. \ref{tab:corr} in the SI.  }
\label{tab:model}
\end{table}



In conclusion, wearables and fitness trackers have been proven to be versatile tools to study physical and health related attributes of individuals in a non-invasive manner. The Oura rings used in the present study enabled us to collect objective health data from participants in realistic settings, thereby providing a unique long-term perspective to investigate the chronotypes of individuals. A major strength of the present study is the finding that even though ETs are most susceptible to health related issues due to misalignment of circadian rhythms, they can improve their health habits over time to some extent. 
Our findings suggest that future research should utilise these non-invasive devices to explore novel and timely interventions through the associated applications with a focus on enhancing health habits as well as maintaining them.


\section*{Materials and methods}
\subsection*{Study design}
Participants were recruited through public advertisements provided online on the university 
website, 
online marketplace and through the student chat groups. A total of nineteen volunteers (nine females and ten males) were chosen for the study between the ages of 24 and 74 and having an average body mass index (BMI) of 24.82 at the start of the study. All participants were healthy and most of them 
worked five days per week. Participants of the study were provided with a detailed information sheet about the study, privacy notice along with consent forms at the beginning of the study. Data collection period lasted from the beginning of October 2023 till the end of September 2024. Data were collected in the form of digital surveys carried out from January 2024 till the end of the study (a total of nine questionnaires) along with the digital health data from Oura rings that participants wore throughout the study period. Sizing kits were provided to obtain the correct sized ring for each of the participants. Health data collected from the rings were stored in the company's cloud through which the researcher accessed the data using APIs. Each participant was associated with a unique identifier before storing in secure Aalto servers and any personal direct identifiers were removed from the data after the data collection period was over. 

\subsection*{Ethics statement}
This study design was reviewed and approved by the Aalto University Research Ethics Committee (approval id: D/134/03.04/2023).

\subsection*{Data}
\begin{figure}[h]
    \centering
    \includegraphics[width=0.7\linewidth]{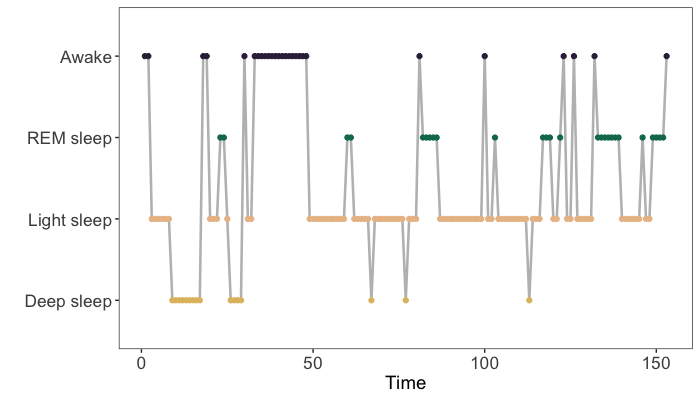}
    \caption{\textbf{Example of a sleep hypnogram provided by Oura ring.} The different stages of sleep recorded during night time sleep is displayed. Each time point in the x-axis represents an interval of five minutes. The four stages of sleep are shown in the y-axis. }
    \label{fig:sleep_hypnogram}
\end{figure}

\begin{figure}[h]
    \centering
    \includegraphics[width=\linewidth]{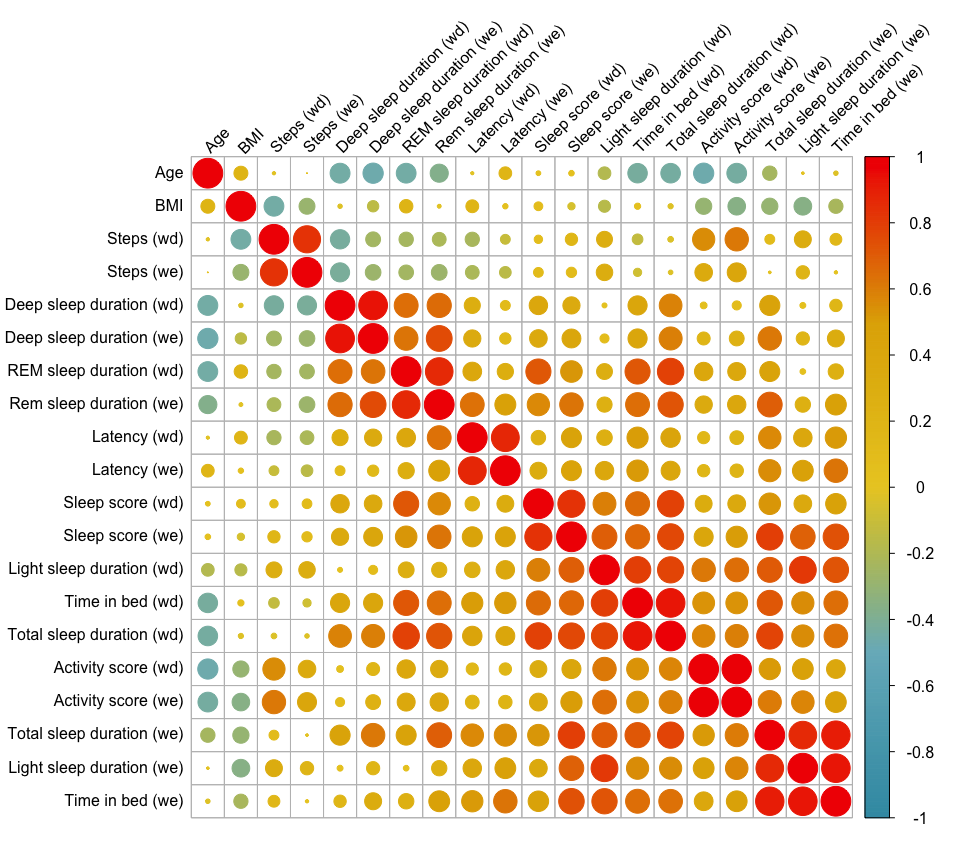}
    \caption{\textbf{Correlation plots of the variables from used in the study along with the age and body mass index (BMI) of the volunteers.} The sleep metrics are mostly correlated with each other and so we chose only one variable (time in bed) as a covariate for the model. }
    \label{fig:corrplot}
\end{figure}

From Oura rings, we collected the bedtimes and awake times of the participants along with night sleep metrics such as the duration of sleep stages and sleep scores. The ring specifies the durations of the different sleep stages within a sleep event: deep, light, rapid eye movement (REM), and awake; and it also provides a hypnogram that classifies the sleep stages every five minutes. An example of the sleep stages within one night sleep event is shown in Fig. \ref{fig:sleep_hypnogram}. It also assigns a sleep score to each sleep event taking into account several sleep-related metrics. For the present study, we have included the time spent in bed during a sleep event and sleep latency measured by the ring. In addition, the ring automatically detects activities such as walking, running, and cycling and the number of steps taken per day by the individual. It assigns an activity score based on several activity-related metrics. Each of the metrics for both the sleep score and the activity score have different weights, and since we have not found any public description of the algorithms for their scoring methods, we consider the score data provided by the ring for the purpose of this paper. 
Definitions of all the variables used in the present paper have been provided in the SI (see Table \ref{tab:variables}). The correlations between the variables along with age and BMI are shown in Fig. \ref{fig:corrplot}. 

Along with objective data collected from the rings, the participants were required to answer the standard morningness-eveningness questionnaire \cite{horne1976self} developed by Horne and Ostberg to get their chronotype and it resulted in seven morning-types (MT), eight neither-types (NT) and four evening-types (ET). A total of nine surveys were conducted within the study period with a minimum of two weeks gap between two consecutive questionnaires to obtain subjective perspective of the participants on their stress levels at different time points. For the present paper, we tracked the stress of the participants through the question ``Stress refers to a situation in which a person feels tense, restless, nervous or anxious or has difficulty sleeping when things are constantly bothering his mind. Think about the last 2 weeks. Have you felt this kind of stress during that time?'' with the options being ``Not at all'', ``Just a little'', ``Somewhat'', ``Quite a lot'', and ``Very much'' which were converted to continuous numerical scale for modelling purposes with 1 being not stressed at all and 5 being very stressed.

\subsection*{Data preprocessing}
For the sleep data, the nights spent in Finland by the participants were considered for analysis. In addition, sleep data of nights during public national holidays of the country was removed from this analysis. This was done to eliminate irregular sleep schedules that may happen due to vacations and holiday celebrations. 
We opted for night-time sleep events (i.e. sleep that starts from 20:00 hrs or later) that lasted 
at least 4 hours and were labelled by Oura as a ``long sleep''.  If a long sleep event occurred during the day, then they were not considered for this analysis. 
This resulted in a total of 5,637 nights for all 19 participants in the study with the mean number of nights per participant being 296.7 and 70.6 as the standard deviation. For the activity data, we removed only Finnish public holidays from our analysis and were left with the total of 6,182 days with the mean number of days per participant being 325.3 and 62.8 being the standard deviation. 
To assess the effects of sleep and activity on the stress of participants we took the averages of activity and sleep metrics fourteen days before the participant answered the stress-related question. For this purpose, in the statistical analysis, we included night sleep and the steps taken during the holidays as well.
\subsection*{Methods}
\subsubsection*{Sleep diversity index:}
An individual's sleep diversity index (SDI) is defined as the percentage of total sleep time that was spent moving between different stages of sleep \cite{suppini2009sleep}. A previous study has shown that SDI is positively correlated with the sleep fragmentation index and negatively correlated with efficiency. It is calculated from an individual sleep hypnogram as shown in Fig. \ref{fig:sleep_hypnogram}. The Oura ring can detect four stages of sleep (deep, light, REM and awake) and in each of the sleep sessions of an individual. Each data point constitutes five minutes of sleep time.

We consider a sampling frequency of 40 minutes that allows for a maximum of 8 data points within each ``analysis window" and calculate the Shannon entropy in them as described in Ref. \cite{suppini2009sleep} using the following equations:
\begin{equation}
    H' = -\sum _{i=1}^S \frac{n_i}{n} \log _2 \frac{n_i}{n}
    \label{eq:shannon_entropy}
\end{equation}
where $n_i$ represents the number of times a sleep stage occurs within a sampling window and n is the total number of sleep stages. S is the total number of samples in each sleep session. If $n_i$ out of $n$
for all $i$, is in the awake stage, then $H'$ = max($H'$). The sleep diversity index is then calculated by the following formula:
\begin{equation}
    SDI = \frac{\sum _{(j>H'_{max}/2)}^{H'_{max}}H'_j} {\sum _{j=1}^N H'_j} X 100
     \label{eq:sdi}
\end{equation}
where $H_j$ is the entropy of the sample $j$. Higher values of SDI mean more fluctuations within a sleep session. 

\subsubsection*{Model}
\label{sec:model}
Since we have repeated measures of each individual, 
we have used the random intercept linear mixed effects (LME) model \cite{gelman2007data} in R using the packages ``lme4" and ``lmeR" to model longitudinal stress\cite{bates2015package}. The participants answered the questions on vastly different days. Therefore, we considered the averages of the time spent in bed on weekdays and weekends separately, fourteen days before the date each of them answered the surveys. This method ensured that the latest objective measurements of sleep were considered as predictors of perceived stress. Additionally, if we did not find data for the fourteen days before the survey answer date, we replaced it by averaging the metric over the entire month. For each individual, we analysed data across nine time points using the random intercept model. The survey answer time, chronotype, and time in bed were included as fixed covariates, while random effects were incorporated to allow each individual to have their own intercept. We have used the following models:

\begin{align}
    \textbf{Null Model:} \quad & y_{it} = \beta_0 + u_{0i} + \epsilon_{it} \label{eq:null_model} \\
    \textbf{Intermediate Model:} \quad & y_{it} = \beta_0 + \beta_1 Z_1(Time) + \beta_2 Z_2(Time) + u_{0i} + \epsilon_{it} \label{eq:intermediate_model} \\
    \textbf{Final Model:} \quad & y_{it} = \beta_0 + \beta_1 Z_1(Time) + \beta_2 Z_2(Time) + \beta_3 Type \notag \\
    & \quad + \beta_4 TIB_{wd} + \beta_5 TIB_{we} + u_{0i} + \epsilon_{it} \label{eq:final_model}
\end{align}

where $y_{it}$ is the perceived stress for each individual $i$ at time point $t~ \epsilon ~\{1,9\}$ modelled as a dependent variables on a continuous scale. $Type$ represents the chronotype of the individual, which is a categorical variable (MT, NT, and ET), and $TIB_{wd/we}$ stands for the time spent in bed on a weekday or a weekend day, respectively, detected by the ring during a sleep event. $Z_1(Time)$ and $Z_2(Time)$ are transformations of $Time$ and $Time^2$ in an orthogonal way to 
reduce collinearity and represent independent linear and quadratic trends, respectively, in the data. $u_{0i}$ is the random intercept for each individual and $\epsilon_{it}$ is the residual error term. $\beta_0$ is the overall mean stress level, $\beta_1$ and $\beta_2$ capture the linear and non-linear time effects, $\beta_3$ models the group effect of chronotypes, and finally,  $\beta_4$ and $\beta_5$ capture the effects of time spent in bed during weekdays and weekends, respectively.


\section*{Supporting information}

\begin{table}[h]
\centering
    \renewcommand{\arraystretch}{1.2}
\begin{tabular}{|l|l|}

\hline
\textbf{Variable }            & \textbf{Definition}                                                                          \\
\hline
\hline
Bedtime start        & Start (date and time) of night sleep event                                          \\
\hline
Bedtime end          & End (date and time) of night sleep event                                            \\
\hline
Sleep score          & Daily sleep score (an integer between 1 and 100)                                    \\
\hline
Deep sleep duration  & Duration of deep sleep within a sleep event in seconds                              \\
\hline
Light sleep duration & Duration of light sleep within a sleep event in seconds                             \\
\hline
REM sleep duration   & Duration of REM sleep within a sleep event in seconds                               \\
\hline
Total sleep duration & Total sleep duration within a sleep event in seconds                                \\
\hline
Sleep latency        & Time taken to fall asleep after going to bed in seconds                             \\
\hline
Time in bed duration & Duration of time spent in bed within a sleep event                                  \\
\hline
Activity score       & Daily activity score (an integer between 1 and 100)                                 \\
\hline
Steps                & Total number of steps taken in a day    \\
\hline

\end{tabular}
\caption{\textbf{Definitions of variables used in the present study.} The sleep score and the activity score are assigned by the Oura company utilising a variety of sleep and activity metrics which can be found in their website. 
}
\label{tab:variables}
\end{table}

\begin{table}[ht!]
\centering
\resizebox{\linewidth}{!}{%
\begin{tabular}{|c|ccccc|}
\hline
 & $Z_1(\text{Time})$ & $Z_2(\text{Time})$ & Neither type & Evening type & Time in bed (wd)  \\
 \hline
 \hline

$Z_1(\text{Time})$ &  & &  &  &  \\

$Z_2(\text{Time})$ & 0.00 &      &      &      &         \\

Neither type        & -0.02 &   0.01   &       &    &           \\

Evening type        & -0.01 &  0.06    &   0.43    &       &           \\

Time in bed (wd)    & -0.08 &    0.06  &    0.10   & 0.15      &             \\

Time in bed (we)    & 0.09 &   -0.05   &    -0.12   & 0.05      &        -0.27     \\
\hline
\end{tabular}%
}
\caption{\textbf{Correlations between the covariates used in the Final model.}  }
\label{tab:corr}
\end{table}

\section*{Acknowledgments}
C.R., and K.K. acknowledge support from EU HORIZON 2020 INFRAIA-1-2014-2015 program project (SoBigData) No. 654024, INFRAIA-2019-1 (SoBigData++) No. 871042 and Brain health through digitalization. 

\bibliography{references}

\end{document}